 \documentclass[3p,twocolumn]{elsarticle}

\usepackage{mathrsfs}
\usepackage{graphicx,epsfig}
\usepackage{epstopdf}
\usepackage{rotating}
\usepackage{subfigure}
\usepackage{color}
\usepackage{amsfonts}
\usepackage{tikz}
\usepackage{amsmath}
\usepackage{amssymb}
\usepackage{hyperref}
\usepackage{relsize}

\begin{document}

\title{Dynamics of skyrmions and edge states in the resistive regime of mesoscopic \emph{p}-wave superconductors}

\author{V.\ Fern\'andez Becerra}
\author{M.\ V.\ Milo\v{s}evi\'c}
\address{Departement Fysica, Universiteit Antwerpen,
Groenenborgerlaan 171, B-2020 Antwerpen, Belgium}
%\affiliation{UNESP-Universidade Estadual Paulista,
%IPMet-Instituto de Pesquisas Metereol\'ogicas, CEP 17048-699 Bauru,
%SP, Brazil}

%\affiliation{Departement Fysica, Universiteit Antwerpen,
%Groenenborgerlaan 171, B-2020 Antwerpen, Belgium}

%\email{Milorad.Milosevic@uantwerpen.be}
%\affiliation{Departement Fysica, Universiteit Antwerpen,
%Groenenborgerlaan 171, B-2020 Antwerpen, Belgium}

\date{\today}

\begin{abstract}
In a mesoscopic sample of a chiral $p\,$-wave superconductor, 
novel states comprising skyrmions and edge states  
have been stabilized in out-of-plane applied magnetic field.
Using the time-dependent Ginzburg-Landau equations we shed light
on the dynamic response of such states to an external applied
current. Three different regimes are obtained, namely, the 
superconducting (stationary), resistive (non-stationary) and 
normal regime, similarly to conventional $s$-wave superconductors.
However, in the resistive regime and depending on 
the external current, we found that moving skyrmions and the edge state behave
distinctly different from the conventional kinematic vortex, thereby providing 
new fingerprints for identification of $p\,$-wave superconductivity.

\end{abstract}

\maketitle

\section{Introduction}

Edge states, appearing where the condensate homogeneity is broken, 
and domain walls, separating regions with different chiralities, 
are the main characteristics of chiral $p\,$-wave superconductivity \cite{JPSJ_matsumoto,PRB_furusaki}. 
They arise as a consequence of breaking the time-reversal symmetry 
in an order parameter with two components, i.e. $\vec{\Psi}=(\psi_+,\psi_-)^T$ \cite{RMP_sigrist}.
Besides the edge states and the domain walls another topological
entity (the skyrmion) has recently emerged in chiral $p\,$-wave superconductivity \cite{SCR_garaud}. 
Unlike the Abrikosov vortex that has a core due to the discontinuity 
of its phase, the skyrmion is coreless and defined by a loop domain wall \cite{PRB_fernandez}. 

Chiral $p\,$-wave superconductivity is realized in spin-triplet 
superconductors. In such materials two electrons pair up 
forming a triplet rather than a singlet as in conventional superconductivity.
In order to fulfil the Pauli principle, the orbital part in spin-triplet
superconductors has odd parity, i.e. angular momentum $L=1$ ($p\,$-wave). 
As a consequence of the spin of the electronic pairs, another topological
entity, the half-quantum vortex (HQV), arises in these materials. 
HQVs are expected to be unscreened by the Meissner effect due 
to their spin currents, i.e. they are likely to be found at the lateral 
borders of the sample \cite{PRL_chung}.

Substantial evidence has been provided over the years that strontium ruthenate, 
Sr$_2$RuO$_4$ (SRO), is a chiral $p\,$-wave superconductor 
\cite{Nat_luke,PRL_kapitulnik,Sci_liu}. 
However, the lack of direct observation of states carrying spontaneous currents
around space homogeneities undermines the candidacy of SRO to the 
$p\,$-wave class of superconducting materials
\cite{PRB_kirtley,PRB_hicks}.
In this work we study the electrical response of skyrmions and edges states
of a mesoscopic chiral $p\,$-wave superconductor sample when an external current 
is applied to the sample. Three different regimes 
are expected, as in conventional superconductivity, namely
superconducting, resistive and normal regime. However, 
the temporal evolution of the two-component superconducting order parameter
($\vec{\Psi}$) is found to provide rich physics,
and depending on the magnitude of applied-current, 
the skyrmionic and edge states must present different behavior 
from kinematic vortices in conventional superconductors
\cite{PHY_andronov,PRL_kinematic,PRB_berdiyorov}.
This in turn provides new possibilities for resistive stages
in the sample behavior, and indirect means to identify
$p\,$-wave superconductivity.

\section{\label{analytical}Theoretical Formalism}

Within the weak-coupling limit and considering a cylindrical Fermi
surface, the dimensionless time-dependent Ginzburg-Landau (TDGL) equations
for the two component order parameter $\vec{\Psi}\!=\!(\psi_+,\psi_-)^T$ and the 
vector potential $\vec{A}\,$, in chiral $p\,$-wave superconductors 
reads
\begin{eqnarray}
\label{fstGL}  
\Bigl(\frac{\partial }{\partial t}\!+\!i\varphi\Bigr)\vec{\Psi}=
\frac{2}{3}\!\left[
 \begin{array}{cc}
  \vec{D}^2  & \Pi_+^2 \\
  \Pi_-^2 & \vec{D}^2 
 \end{array}
\right]
\left(
\begin{array}{c}
\psi_+ \\
\psi_-
\end{array}
\right) 
\nonumber\\
+ \vec{\Psi}\Bigl(1-\frac{1+\tau}{2}|\vec{\Psi}|^2 \pm\frac{\tau}{2}\vec{\Psi}^*\hat{\sigma}_z\vec{\Psi}\,\Bigl) \, ,\\
\kappa^2\nabla\times(\nabla\times\vec{A}) + \Bigl(\nabla\varphi+\frac{\partial\vec{A}}{\partial t}\Bigr)= \vec{J}_s\, , 
\label{sndGL}  
 \end{eqnarray}

\noindent where $\varphi$ is the electrostatic potential, 
$\vec{D}=(\vec{\nabla}-i\vec{A})$ is the covariant derivative,
$\Pi_{+(-)}$ is the Landau level creation (annihilation) operator, 
$\tau$ is a phenomenological parameter, $\hat{\sigma}_z$ 
is a Pauli matrix, $\kappa$ is the GL parameter, 
and $\vec{J}_s$ is the superconducting current density,
\begin{eqnarray}
 \vec{J}_s &=&\! \frac{1}{3}\,{\rm Im}\Bigl\{\psi_+^*\vec{D}\psi_+ + \psi_-^*\vec{D}\psi_- \Bigr\} \nonumber\\ 
&+&\! \frac{1}{3\sqrt{2}}\,{\rm Im}\Bigl\{\vec{\Psi}^*\Bigl[\Pi_+\hat{\sigma}_+ \!+\! \Pi_-\hat{\sigma}_-\Bigr]\vec{\Psi}\,\hat{\imath}\nonumber\\
&+&\! i\,\vec{\Psi}^*\Bigl[\Pi_+\hat{\sigma}_+ \!-\! \Pi_-\hat{\sigma}_-\Bigr]\vec{\Psi}\,\hat{\jmath} \Bigr\},  
\label{sucurr}
\end{eqnarray}

\noindent where $\hat{\sigma}_\pm=(\hat{\sigma}_x\pm i\hat{\sigma}_y)/2$,
and $\{\hat{\imath},\hat{\jmath}\}$ is the canonical base in Cartesian coordinates.
In Eqs. (\ref{fstGL})-(\ref{sucurr}) distances are scaled to the 
superconducting coherence length $\xi$, 
time to the GL time $t_0$,  and
the vector and electrostatic potentials to $A_0=\hbar c/2e\xi$ 
and $\varphi_0=A_0/ct_0$, respectively. Similarly,
the order parameter is scaled to its bulk zero-field value $|\vec{\Psi}(\vec{A}\!=\!0)|$,
and the current density to $J_0=(e\hbar/m\xi)|\vec{\Psi}_0|^2$.
In order to study the dynamical properties of mesoscopic chiral $p\,$-wave 
superconductors, we adopt the Coulomb gauge, i.e. $\vec{A}$ is divergence-free at all times, 
since it provides the equation for the electrostatic potential,
\begin{equation}
\centering
 \nabla^2\varphi = \nabla\!\cdot\!\vec{J}_s\,.
\label{poteqn}
 \end{equation}

For the vector potential and because of an out-of-plane applied magnetic field, 
we choose $\vec{A}\!=\!-(\vec{r}\!\times\!\vec{H})/2$.
The boundary conditions imposed at the superconductor-vacuum and 
superconductor-normal-metal interfaces are, 
\begin{eqnarray}
\left .\
\begin{array}{r}
\vec{\Psi} = 0 \\
\partial_y\varphi + j=0 
\end{array}
\!\right\}
\text{  at N and S sides},& 
\nonumber \\
\left.\
\begin{array}{r}
\psi_+\! +\! \psi_- \!=\! 0 \\
D_x\psi_+\!-\! D_x\psi_- \!=\! 0\\
\partial_x\varphi \!=\! 0
\end{array}
\!\right\}
\text{  at E and W sides,}
\label{bcond}
 \end{eqnarray}

\noindent respectively. N, S, E and W stand for the cardinal points.
Current $j$ is applied at contacts located at N and S. 
Eq. (\ref{bcond}) completes the TDGL equations for chiral $p\,$-wave superconductors
which we solve using the finite-difference technique. 

\section{\label{Results}Results}
%\subsection{}
%\subsubsection{}

In this work we stabilize skyrmions and edge states 
in a mesoscopic sample of size $10\xi\!\times\!12\xi$ by applying 
an external magnetic field $H\!=\!0.8H_{c2}$ out of plane of the sample. 
Although skyrmions can be stabilized also in bulk samples \cite{SCR_garaud}, 
the edge states containing one vortex just in component $\psi_+$,
i.e. a half-quantum vortex for the system, appear only where
the space homogeneity is broken, thus are characteristic
of mesoscopic samples \cite{PRB_fernandez}. 
In what follows, we examine the response of such states
to applied current. In our study, the external current 
density $j$ is increased adiabatically from zero up to certain value 
$j_f$, streaming from the north to the south side of the sample.

\begin{figure}[hhtt]
\center
\includegraphics[clip=true,trim={0.0cm 0.05cm 0.0cm 0.0cm}, scale=0.7]{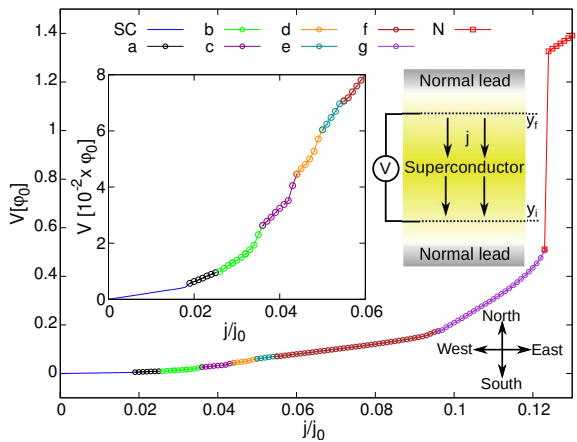}
\caption{Measured voltage versus applied current in dimensionless units for 
a mesoscopic sample of size $10\xi\!\times\!12\xi$ with normal 
contacts at north and south sides. An external magnetic field
($H\!=\!0.8H_{c2}$) applied perpendicularly to the sample
stabilizes domain walls, skyrmions and half-quantum vortices. Three different
regimes can be identified, namely the superconducting (SC), 
resistive and normal regime (N). The resistive regime contains
different states labelled here by ($a$) - ($g$).}
\label{IVplt}
\end{figure}

The plot of voltage against current for a mesoscopic chiral 
$p\,$-wave superconductor is shown in Fig. \ref{IVplt}, with the voltage
defined as: $V = {\bar\varphi}|_{y_i}\!-{\bar\varphi}|_{y_f}$, 
where the bar over the electrostatic potential denotes average, and $y_i=1.5\,\xi$
and $y_f=10.5\,\xi$. To date, for chiral $p\,$-wave superconductors
only the stationary GL equations have been derived either phenomenologically
or microscopically \cite{RMP_sigrist,PRB_furusaki,PRB_zhu}. 
The TDGL equations (\ref{fstGL}) and (\ref{sndGL}), obtained as an 
extension of the stationary ones after imposing full gauge invariance
are conceived for gapless superconductors, but  
are expected to capture the evolution of static and dynamic states
in the here studied cases.

\begin{figure}[ttt]
\includegraphics[trim={1.0cm 0.6cm 0.34cm 1.2cm}, clip=true, scale=0.625]{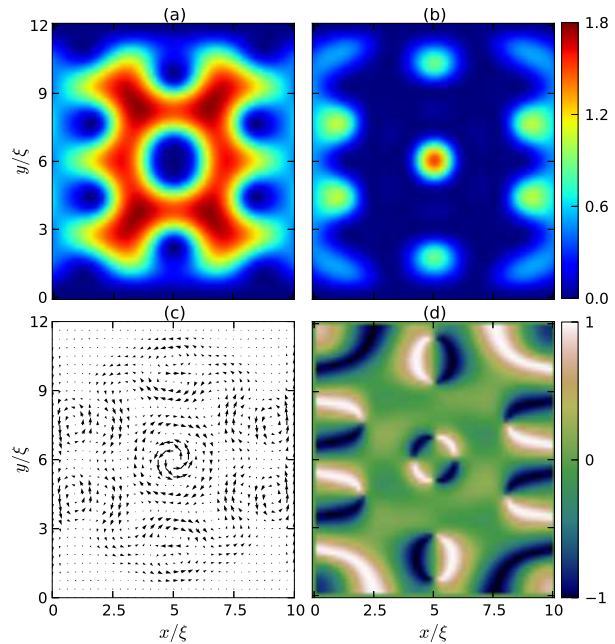}
\caption{Contour plots of $|\psi_+|^2$ (a), $|\psi_-|^2$ (b), 
the current distribution $\vec{J}_s$ (c) and the intercomponent
phase difference $\cos{(\theta_x\!-\!\theta_y)}$ (d).
Precisely, the phase difference $(\theta_x\!-\!\theta_y)$ is obtained from the 
fields: $\psi_x=(\psi_+\!+\! \psi_-)/2$ and $\psi_y=(\psi_+\!-\!\psi_-)/2i$.
According to panel (d) the superconducting state is composed of 
one skyrmion inside the sample and the edge state 
containing six half-quantum vortices at the borders and four domain walls
around the corners.}
\label{statstate}
\end{figure}

Three different regimes can be identified from the 
current-voltage characteristics of Fig. \ref{IVplt}, 
namely the superconducting (stationary), resistive (non-stationary), 
and normal (ohmic) regime. At low currents the 
superconducting regime can exhibit weak resistance,
consequence of the normal contacts (see the inset of Fig. \ref{IVplt}). 
Fig. \ref{statstate} shows the superconducting phase at $j=0$ in contour plots
of $|\psi_+|^2$ (a), $|\psi_-|^2$ (b), the current density $\vec{J}_s$ (c),  
and the cosine of the intercomponent phase difference $\cos{(\theta_x\!-\!\theta_y)}$ (d)
[from now on simply call the phase difference]. 
The angular phases $\theta_x$ and $\theta_y$ are   
obtained from the redefined order parameters 
$\psi_x=(\psi_+\!+\! \psi_-)/2$ and $\psi_y=(\psi_+\!-\!\psi_-)/2i$, respectively.
The superconducting state, according to panel (d) 
is composed of one skyrmion inside the sample and 
the edge state enclosing the sample and containing 
six HQVs at the borders, and four domain walls around the corners. 

\begin{figure}[ttt]
\includegraphics[trim={0.8cm 0.2cm 1.2cm 0.97cm}, clip=true, scale=0.91]{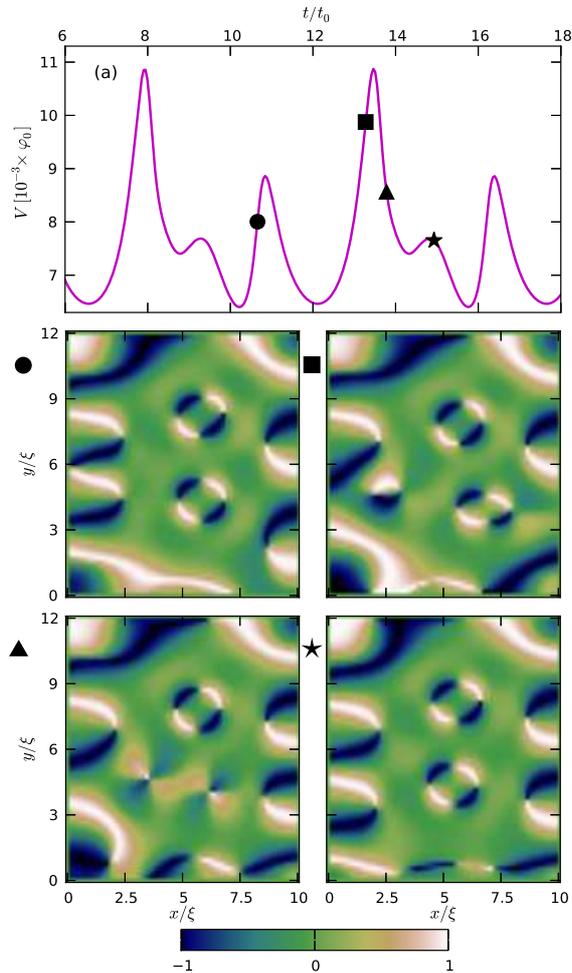}
 \caption{Temporal evolution of the non-stationary state $a$ 
 of Fig. \ref{IVplt} ($j\!=\!0.022j_0$), seen in the voltage vs.
 time plot (a), and four snapshots of the phase 
 difference $\cos{(\theta_x\!-\!\theta_y)}$, 
($\mathlarger{\mathlarger{\bullet}}$), 
({$\mathsmaller{\blacksquare}$}), ($\blacktriangle$) and ($\mathlarger{\mathlarger{\star}}$), respectively. 
The snapshots describe the states asociated with the three peaks clearly seen in the voltage plot (a).}
 \label{fstnonst}
\end{figure}  

As one increases the external current the superconducting state of Fig. \ref{statstate}
shifts to the right due to the reduction of the superconducting currents in the east side 
compared to the west side (due to compensation of the Meissner currents with applied current,
see e.g. \cite{PRL_berdiyorov}). The resistive regime thus appears at currents  
where the flux motion drives the superconductor to a non-stationary state. 
From Fig. \ref{IVplt} one can see that such regime 
exhibits sequential jumps in the voltage 
as current is increased, which we attribute
to different non-stationary states (labelled there by letters). 
In order to study the temporal evolution of 
the two-component superconducting order parameter 
in the resistive regime, we choose the state $a$ of Fig. \ref{IVplt} 
since it summarizes all the rich properties that a mesoscopic chiral 
$p\,$-wave superconductor presents. Animated data of the remaining 
states of Fig. \ref{IVplt} are therefore left for the supplementary section ($a$ included). 

The plot of Fig. \ref{fstnonst}(a) reveals that the 
voltage in state $a$ of Fig. \ref{IVplt} is a periodic 
function of time. Moreover, one can clearly see 
that there exist three distinct modes that 
correspond to a special flux motion. Contour plots of 
the phase difference show the superconducting state at 
these modes. From panel ($\mathlarger{\mathlarger{\bullet}}$) to 
($\mathsmaller{\blacksquare}$) 
one can distinguish three events: (i) the bottom skyrmion 
is heading towards the E side, (ii) one HQV at the E side
left the sample at the south-east corner, 
and (iii) one HQV at the W side acquired a quantum 
of flux from component $\psi_-$ to form 
a full vortex. Next, the skyrmion having two quanta
of flux broke into two HQVs and one of these went 
to the E side while the other fused with another quantum 
of flux to form a second full vortex [see panels 
($\mathsmaller{\blacksquare}$) and ($\blacktriangle$)].
Another mechanism of skyrmion annihilation, displayed 
in state $f$ (see the supplementary section), consist of
one skyrmion losing its two quanta of flux in the form 
of two concentric HQVs. Finally, panels ($\blacktriangle$)
to ($\mathlarger{\mathlarger{\star}}$) show 
the fusion of two full vortices into a skyrmion and the 
nucleation of a HQV at the W side from the west-south corner.
There exists another mechanism of skyrmion creation, 
consisting of two quanta of flux being pumped inside the sample
from the edge state, more precisely the W side (see the state $f$ in  
the supplementary section).

%\begin{figure}[ttt]
%\includegraphics[trim={3.4cm 0.8cm 1.8cm 1.05cm}, clip=true, scale=0.88]{four_restwo.eps}
% \caption{Dynamical evolution of the non-stationary state $f$ 
% of Fig. \ref{IVplt}, seen in the voltage vs.
% time plot (a), and three contour plots of the phase 
% difference $\cos{(\theta_x\!-\!\theta_y)}$, 
%($\ominus$), ($\square$) and ($\triangle$), respectively. 
%Two fast modes (peaks) with different amplitudes and modulated 
%by a slow mode are seen in the voltage plot (a). 
%The circle, square and triangle markers indicate the two 
%fast modes.}
% \label{sndnonst}
% \end{figure}

The role of the normal contacts in the 
one dimensional movement of the HQVs is crucial.
Owing to the superconducting-normal-metal interfaces 
the barrier for HQV exit/entry is cancelled
on the N/S sides of the sample.
Further, there also exists a barrier  
formed by Meissner currents on the E/W sides,
that prevents the HQVs to leave the edge state or conversely 
that prevent the HQV to get in the sample.
Altogether, the HQV at the E and W sides 
experience the easy direction for motion along  
the superconducting-vacuum interfaces. 

\section{\label{summary}Conclusions}

In summary, using the time-dependent Ginzburg-Landau equations
for chiral $p\,$-wave superconductors, we have shown 
some characteristic dynamics of 
skyrmions and the edge state in a mesoscopic 
$p\,$-wave superconductor. When an external current
is applied to the sample, the resistive state shows
much richer behavior compared to conventional $s$-wave superconductors.
For example, depending on the strength of the external current, 
we found that the half-quantum vortices in the edge state can
move along the direction of the applied current, 
contrary to standard kinematic vortices
which always move perpendicularly to the current flow 
\cite{PHY_andronov, PRL_kinematic, PRB_berdiyorov}.
We also observe in the resistive regime that under the applied current
skyrmions either nucleate the sample directly from the edge state 
or arise from the recombination of two full vortices.
These findings combinatorially increase 
the possibilities for different resistive states in mesoscopic
superconductors, worthy of further exploration.

% Surround figure environment with turnpage environment for landscape
% figure
% \begin{figure}
% \end{figure}

% tables should appear as floats within the text
% The ruledtabular enviroment adds doubled rules to table and sets a
% reasonable default table settings.
% Use the table* environment to get a full-width table in two-column
% Add \usepackage{longtable} and the longtable (or longtable*}
% environment for nicely formatted long tables. Or use the the [H]
% placement option to break a long table (with less control than
% \begin{table}%[H] add [H] placement to break table across pages
% \begin{ruledtabular}
% Lines of table here ending with \\
% \end{tabular}
% \end{ruledtabular}
% \end{table}

% Surround table environment with turnpage environment for landscape
% table
% \begin{table}
% \begin{ruledtabular}
% \end{tabular}
% \end{ruledtabular}
% \end{table}

%\appendix
%\section{}

%\begin{acknowledgments}
\section{Acknowledgments}
This work was supported by the Research Foundation - Flanders (FWO),
and COST MPNS Action MP-1201, FP7.
%\end{acknowledgments}

\section*{References}
\bibliography{bibfile_physC}

\end{document}